\documentclass[useAMS,usenatbib]{mn2e}
\usepackage{graphicx}
\title{A new method to constrain annihilating dark matter}
\author[ Chan \& Lee]{ Man Ho Chan \thanks{chanmh@eduhk.hk} and Chak Man Lee
\\ Department of Science and Environmental Studies, The Education University of Hong Kong, Tai Po, Hong Kong}

\begin{document}

\date{Accepted XXXX, Received XXXX}

\pagerange{\pageref{firstpage}--\pageref{lastpage}} \pubyear{XXXX}

\maketitle

\label{firstpage}

\date{\today}

\begin{abstract}
Recent indirect searches of dark matter using gamma-ray, radio, and cosmic-ray data have provided some stringent constraints on annihilating dark matter. In this article, we propose a new indirect method to constrain annihilating dark matter. By using the data of the G2 cloud near the Galactic supermassive black hole Sgr A*, we can get stringent constraints on the parameter space of dark matter mass and the annihilation cross section, especially for the non-leptophilic annihilation channels $b\bar{b}$ and $W^{\pm}$. For the thermal annihilation cross section, the lower bounds of dark matter mass can be constrained up to TeV order for the non-leptophilic channels with the standard spike index $\gamma_{\rm sp}=7/3$. 
\end{abstract}


\section{Introduction}
The nature of dark matter is one of the most important mysteries in astronomy and astrophysics. Dark matter particles are almost collisionless and they nearly do not interact with ordinary matter. Therefore, it is very difficult to observe dark matter particles directly. In the past few decades, astrophysicists started to observe and constrain the potential indirect signals from dark matter \citep{Ackermann,Egorov,Storm,Ackermann2,Albert,Ambrosi,Boddy,Chan,Aguilar,Albert2,Li,Abdalla,Chan2,Chan3,Chan4,Acciari}. Some theories predict that dark matter particles can self annihilate to give high energy particles such as electron-positron pairs, photons, and neutrinos \citep{Profumo}. Therefore, indirect-detection of dark matter has become a popular way to investigate the nature of dark matter.

There are two parameters involved in the indirect-detection of dark matter: dark matter mass $m_{\rm DM}$ and the annihilation cross section $\langle \sigma v \rangle$. Dark matter particles would undergo self-annihilation via the standard model channels (e.g. $e^{\pm}$ channel or $b{\bar b}$ quark channel). The amount of the high energy particles produced in the dark matter self-annihilation process can be predicted even though we do not know the nature of dark matter \citep{Cirelli,Profumo}. Many previous studies using the observational data of cosmic rays \citep{Ambrosi,Aguilar}, gamma rays \citep{Ackermann,Ackermann2,Albert,Boddy,Abazajian,Li,Abdalla,Acciari}, neutrinos \citep{Albert2,Chan3}, X-ray flux \citep{Chan4} or radio flux \citep{Egorov,Storm,Chan,Regis2,Chan2} have obtained some stringent constraints of dark matter parameters. For the value of thermal annihilation cross section predicted by standard cosmology $\langle \sigma v \rangle=2.2 \times 10^{-26}$ cm$^3$/s \citep{Steigman}, many recent results favour $m_{\rm DM} \ge 100$ GeV for annihilating dark matter \citep{Chan,Abazajian,Regis2,Chan2,Chan4}.

In this article, we propose an entirely new way to constrain annihilating dark matter. Observations show that there exists a large cloud, called G2 cloud, orbiting the supermassive black hole at the Galactic Centre (Sgr A*) with a very small pericenter $r_p \approx 267 \pm 20$ AU \citep{Gillessen}. Theoretical models show that the dark matter density can be very high near a supermassive black hole (i.e. the dark matter density spike) if the supermassive black hole grows adiabatically from a smaller seed \citep{Gondolo,Gnedin,Merritt,Sadeghian}. Such a very high dark matter density would greatly enhance the dark matter annihilation rate so that a large amount of high-energy particles would be produced \citep{Gondolo,Gnedin,Bertone,Fields,Shapiro}. In particular, the electrons and positrons produced from dark matter annihilation would heat up the G2 cloud and maintain a certain high temperature. By modelling the heating and cooling rate of the G2 cloud, we can effectively constrain the relevant parameters of dark matter. Following the standard dark matter density spike model \citep{Gondolo,Fields}, we show that this new method can give stringent constraints on dark matter mass and the annihilation cross section.

\section{The theoretical framework}
The orbital motion of the G2 cloud has been monitored for several years \citep{Gillessen,Phifer,Pfuhl,Valencia,Schartmann,Gillessen2}. It has a small semi-major axis $a \approx 0.02$ pc (assumed the distance of the Earth from the Galactic Centre is $D=8.3$ kpc \citep{Gillessen3}) and a very high orbital eccentricity $e=0.9384 \pm 0.0066$ \citep{Gillessen}. Therefore, the distance between the G2 cloud and the Sgr A* at the pericenter is very small $r_p \approx 1.29$ mpc. The G2 cloud passed the pericenter position in 2013 \citep{Gillessen,Schartmann}. 

If dark matter particles are collisionless, theoretical models show that dark matter surrounding a supermassive black hole would re-distribute to form a dark matter density spike if the supermassive black hole grows adiabatically \citep{Gondolo,Gnedin,Merritt}. Some indirect evidence of the dark matter density spike is found for the stellar-mass black hole systems \citep{Chan5}. A standard dark matter density spike model at the Galactic Centre can be expressed as \citep{Fields}
\begin{equation}
\rho_{\rm DM}(r)=\frac{\rho_{\rm sp}(r)\rho_{\rm in}(t,r)}{\rho_{\rm sp}(r)+\rho_{\rm in}(t,r)},\,\,\,\,4GM/c^2 \le r \le r_b\,\,{\rm (spike)}
\label{density}
\end{equation}
with
\begin{eqnarray}
r_{\rm b}&=&0.2r_{\rm h}\\
\rho_{\rm sp}(r)&=&\rho_{\rm b}(r_{\rm b}/r)^{\gamma_{\rm sp}}\label{densityrhosp}\\
\rho_{\rm in}(t,r)&=&\rho_{\rm ann}(t)(r/r_{\rm in})^{-\gamma_{\rm in}}.\label{densityrhoin}
\end{eqnarray}
Here, $r_h$ is the radius of influence of the supermassive black hole and $\gamma_{\rm sp}$ is the spike index. The parameters involved in the spike model can be found in Table 1. The density $\rho_{\rm ann}(t)$ is the so-called DM `annihilation plateau' density $\rho_{\rm ann}(t)=m_{\rm DM}/\langle\sigma v\rangle t$, reached by $\rho_{\rm sp}(r)$ in the innermost region of the spike at $r=r_{\rm in}$ and $t=t_{\rm ann}$ the lifetime over which annihilation have occurred (i.e. the age of the supermassive black hole $t_{\rm ann} \sim 10^{10}$ yr) \citep{Fields}. Here, we take the mass of the supermassive black hole Sgr A* $M=4.15 \times 10^6M_{\odot}$ \citep{Abuter}. Outside the spike region, we assume that the dark matter density follows the Navarro-Frenk-White (NFW) density profile \citep{Fields}:
\begin{equation}
\rho=\rho_{\rm s}\frac{r_{\rm s}}{r},
\end{equation}
where $r_{\rm s}$ and $\rho_{\rm s}$ are the scale radius and scale density of the NFW profile. Therefore, we have $\rho_{\rm b}=\rho_{\rm s}r_{\rm s}/r_{\rm b}$. Recent rotation curve data of the Milky Way indicate $\rho_s= 0.0182\pm 0.0074 \;M_{\odot}{\rm pc}^{-3}$ and $r_s= 10.7\pm 2.9\,\,{\rm kpc}$ \citep{Sofue}. For the NFW profile outside the spike region, we expect that the spike index should be $\gamma_{\rm sp}=7/3$ \citep{Fields}. However, if the stellar heating effect is important, the spike index would be $\gamma_{\rm sp} \approx 1.5$ \citep{Gnedin,Merritt}. These two values of the spike index ($\gamma_{\rm sp}=1.5$ and $\gamma_{\rm sp}=7/3$) will be considered in our analysis. In Fig.~1, we can see that the dark matter density spike profile would follow $\rho_{\rm sp}$ at large $r$ and converge to $\rho_{\rm in}$ at small $r$. The transition of the change in mass density profile depends on the dark matter mass $m_{\rm DM}$ and the annihilation cross section $\langle \sigma v \rangle$, which determine the value of $\rho_{\rm in}$ through $\rho_{\rm ann}$. 

If dark matter particles can self-annihilate, their extremely high density inside the spike region would greatly enhance the rate of dark matter annihilation. A large amount of high-energy electron and positron pairs would be produced during the annihilation process. These electrons and positrons will diffuse out and cool down. The cooling is dominated by four processes: synchrotron, inverse Compton scattering, Bremsstrahlung, and Coulomb loss. The total cooling rate $b_T(E,r)$ (in unit of $10^{-18}$ GeV s$^{-1}$) for these four processes can be expressed as \citep{Colafrancesco,Egorov2}
\begin{eqnarray}
b_T(E,r)&=&2.54E^2B^2 \nonumber\\
&&+ 25E^2 \nonumber\\
&&+ 7.1\times10^{-2}\gamma n(r)\left(0.36+\ln\gamma\right)\nonumber\\
&&+7.6 n(r)\left[73+\ln\left(\frac{\gamma}{n(r)}\right)\right].
\label{cooling}
\end{eqnarray}
where $E$ is the energy (in GeV) of the high-energy electrons or positrons produced, $n(r)$ is the thermal electron number density in cm$^{-3}$, $B$ is the magnetic field strength in $\mu$G, and $\gamma$ is the Lorentz factor of the high-energy electrons or positrons. Note that the inverse Compton scattering cooling term $25E^2$ in Eq.~(6) has assumed the contribution of the Cosmic Microwave Background (CMB) photons only. In fact, the radiation density at the Galactic Centre should be much higher than the CMB radiation density. Nevertheless, since the radiation density at the deep Galactic Centre is unknown and the inverse Compton scattering cooling is nearly negligible compared with the synchrotron cooling, we have taken the conservative CMB radiation density in Eq.~(6) only. At the deep Galactic Centre, the magnetic field strength can be as large as 1 mG \citep{Eatough}. By following the equipartition condition (i.e. magnetic energy balances the kinetic pressure) within the accretion region and the magnetic flux conservation outside the accretion region, the magnetic field strength at the deep Galactic Centre can be expressed by \citep{Aloisio,Regis}
\begin{equation}
B=\left\{\begin{array}{ll}
B_0\left(\frac{r_{\rm c}}{r}\right)^{5/4}& {\rm for}\,\,r\le r_{\rm c},\\
B_0\left(\frac{r_{\rm c}}{r}\right)^{2}& {\rm for}\,\,r> r_{\rm c}.
\end{array}\right.
\end{equation}
where the magnetic field $B_0=7.2$ mG at radius $r_{\rm c}=0.04$ pc. 

Generally speaking, the cooling and diffusion of the high-energy electrons and positrons can be governed by the diffusion-cooling equation \citep{Atoyan}. The cooling rate and diffusion rate can be characterised by the cooling timescale $t_c$ and the diffusion timescale $t_d$ respectively. For the cooling processes, synchrotron cooling is the most dominant one. For $r \le r_p$, the synchrotron cooling rate is $b_s \sim 10^{-4}-10^{-2}$ GeV s$^{-1}$ for $E \sim 10-100$ GeV. Therefore, the cooling timescale is $t_c \sim E/b_s \sim 10^4-10^5$ s. For the diffusion process, the diffusion coefficient is unknown for the region at the Galactic Centre. Nevertheless, \citet{Istomin} show that the diffusion coefficient depends on the Larmor radius $r_L$ of the electrons and positrons for small Larmor radius: $D \approx (\pi/2)v_0r_L$, where $v_0$ is the particle velocity. Since the magnetic field at the Galactic Centre is very high, the Larmor radius is very small $r_L \sim 10^{11}-10^{13}$ cm for $E \sim 10-100$ GeV. By taking $v_0 \sim c$, we get $D \sim 10^{22}-10^{23}$ cm$^2$ s$^{-1}$, which is close to the B\oe hm diffusion regime \citep{Boehm}. Since the diffusion timescale is roughly given by $t_d \sim r_p^2/D$, we get $t_d \sim 10^9-10^{10}$ s. As the cooling timescale is much less than the diffusion timescale, the cooling is much more efficient and the diffusion term in the diffusion-cooling equation can be neglected \citep{Vollmann}. In this regime, the equilibrium energy spectrum of the electrons or positrons is simply related to the injection source from the dark matter annihilation \citep{Egorov,Storm}
\begin{equation}
\frac{dn_{\rm e}}{dE}(E,r)=\frac{\langle \sigma v \rangle [\rho_{\rm DM}(r)]^2}{2m_{\rm DM}^2b_T(E,r)}\int_E^m\frac{dN_{\rm e,inj}}{dE'}dE',
\end{equation}
where $dN_{\rm e,inj}/dE$ is the injection energy spectrum of the dark matter annihilation, which can be predicted by numerical calculations \citep{Cirelli}.

In general, the cooling via Coulomb scattering is collisional while the cooling via synchrotron, inverse Compton scattering and Bremsstrahlung is radiative (e.g. mainly radio, X-ray and gamma-ray emissions). Therefore, the energy of the electrons and positrons produced from dark matter annihilation can transfer to the G2 cloud via collisional Coulomb scattering only. The heating rate per unit volume $Q$ for both positron and electron contribution is then given by
\begin{equation}
Q=2\int_0^{\infty}\frac{dn_{\rm e}}{dE}(E,r)\times b_C(E,r)dE,
\end{equation}
where $b_C(E,r)$ is the Coulomb cooling (heating) rate
\begin{equation}
b_C(E,r)=7.6 n(r)\left[73+\ln\left(\frac{\gamma}{n(r)}\right)\right].
\label{cooling1}
\end{equation}

Consider the G2 cloud at the closest distance from the Sgr A* (i.e. the pericenter). The dark matter annihilation rate would be larger because the dark matter density is higher at the pericenter. Therefore, we expect that the heating rate of the electrons and positrons due to dark matter annihilation $Q$ would be the largest at the pericenter. Observations show that the temperature of the G2 cloud is $T=550 \pm 90$ K \citep{Gillessen}. The temperature did not change after the G2 cloud passed the pericenter position in 2013 \citep{Zajacek}. Therefore, we can assume that the temperature is almost constant $T=550 \pm 90$ K at the pericenter. The cooling rate per unit volume of a gas cloud for $T \sim 500-5000$ K can be well described by \citep{Smith,Wadekar}
\begin{equation}
\Lambda=10^{-27.6} \left[\frac{n(r_p)}{1~\rm cm^{-3}} \right]^2 \left(\frac{T}{1~\rm K} \right)^{0.6}\,\,{\rm erg\, cm^{3}\,s^{-1}}.
\end{equation}
The electron number density of the G2 cloud can be given by \citep{Gillessen}
\begin{equation}
n(r_p)=2.6\times 10^5 f_{\rm v}^{-1/2}R_{\rm c,15\, mas}^{-3/2}{T_{e,4}}^{0.54} {\rm cm^{-3}},
\end{equation}
where $T_{e,4}$ is the electron temperature in the unit of $10^4$ K. Assuming the filling factor $f_{\rm v}=1$, the size of the G2 cloud $R_{\rm c}=15$ mas and a larger electron temperature $T_{e,4}=1$, we take the number density to be $n(r_p) \sim 2.6 \times 10^5$ cm$^{-3}$ for analysis, which is a more conservative value for cooling rate estimation. Therefore, the cooling rate per unit volume of the G2 cloud at the pericenter is $\Lambda=7.48 \times 10^{-16}$ erg cm$^{-3}$ s$^{-1}$. To obtain the constraints of the annihilation cross section $\langle \sigma v \rangle$ and the dark matter mass $m_{\rm DM}$, we set $Q \le \Lambda$.

\section{Results}
In our analysis, we consider two benchmark values of the spike index $\gamma_{\rm sp}=1.5$ and $\gamma_{\rm sp}=7/3$. In Fig.~2, by taking the thermal annihilation cross section predicted by standard cosmology $\langle \sigma v \rangle=2.2 \times 10^{-26}$ cm$^3$ s$^{-1}$ \citep{Steigman}, we plot the heating rate per unit volume for five popular standard model annihilation channels ($e^{\pm}$, $\mu^{\pm}$, $\tau^{\pm}$, $b\bar{b}$ and $W^{\pm}$). For $Q \le \Lambda$, we can see that there exist minimum allowed dark matter mass $m_{\rm DM}$ for some annihilation channels. These values are regarded as the lower bounds of $m_{\rm DM}$ for these channels. We can see that this new analysis can provide very stringent constraints on $m_{\rm DM}$, especially for the non-leptophilic channels (i.e. $b\bar{b}$ and $W^{\pm}$) with $\gamma_{\rm sp}=7/3$ (see Table 2). For the standard spike index $\gamma_{\rm sp}=7/3$, only $m_{\rm DM}>1$ TeV would be favoured for these two channels, which are more stringent than the constraints obtained previously \citep{Albert,Chan,Abazajian,Regis2,Chan4}. However, if stellar heating is significant so that a smaller spike index is resulted $\gamma_{\rm sp}=1.5$, only the $b\bar{b}$ channel can give a meaningful lower bound $m_{\rm DM} \ge 20$ GeV.   

In fact, the value of the annihilation cross section is model-dependent. If the annihilation cross section is velocity-dependent (e.g. p-wave annihilation), its value can be greater than the standard thermal annihilation cross section \citep{Zhao}. Therefore, the annihilation cross section might be larger near the Galactic Centre as the dark matter velocity dispersion is large near the Sgr A*. If we release $\langle \sigma v \rangle$ to be a free parameter, we can obtain forbidden ranges of annihilation cross section for different $m_{\rm DM}$ and annihilation channels (see Fig.~3). In general, from Eqs.~(1)-(4), we can see that the heating rate $Q \propto \langle \sigma v \rangle [\rho_{\rm in}(t,r)]^2 \propto 1/\langle \sigma v \rangle$ for $\langle \sigma v \rangle$ is large while $Q \propto \langle \sigma v \rangle [\rho_{\rm sp}(r)]^2 \propto \langle \sigma v \rangle$ for $\langle \sigma v \rangle$ is small. Therefore, there exists a peak of $Q$ for a certain $\langle \sigma v \rangle$ as shown in Fig.~3. For example, in order to satisfy $Q \le \Lambda$ for the case of $\gamma_{\rm sp}=7/3$, we get $\langle \sigma v \rangle \le 2.5 \times 10^{-27}$ cm$^3$ s$^{-1}$ or $\langle \sigma v \rangle \ge 1.1\times 10^{-22}$ cm$^3$ s$^{-1}$ for $m_{\rm DM}=1000$ GeV annihilating via the $b\bar{b}$ channel. In other words, the range of $\langle \sigma v \rangle=2.5\times 10^{-27}-1.1\times 10^{-22}$ cm$^3$ s$^{-1}$ is forbidden for this particular $m_{\rm DM}$ and annihilation channel. Therefore, in Fig.~4, we plot the forbidden parameter space of $\langle \sigma v \rangle$ and $m_{\rm DM}$ for the $b\bar{b}$ and $W^{\pm}$ channels. A large parameter space of $m_{\rm DM}$ and $\langle \sigma v \rangle$ would be ruled out based on this new analysis.

\section{Discussion}
In this article, we present the analysis of using the data of the G2 cloud near the supermassive black hole Sgr A* to constrain the parameters of annihilating dark matter. By examining the heating rate and cooling rate of the G2 cloud, we can get some stringent constraints on the annihilation cross section and the dark matter mass, especially for the $b\bar{b}$ and $W^{\pm}$ channels. This can provide new constraints on dark matter parameters, which are supplementary to the traditional gamma-ray constraints, radio constraints and cosmic-ray constraints. In particular, for the standard benchmark value of the spike index $\gamma_{\rm sp}=7/3$, the G2 cloud data can constrain the dark matter mass lower bound up to TeV order with the thermal annihilation cross section for the $b\bar{b}$ and $W^{\pm}$ channels. These bounds are much more stringent than the previous bounds obtained using radio \citep{Egorov,Chan,Regis2,Chan2} and gamma-ray data \citep{Albert,Abazajian,Abdalla,Acciari}.

However, our results are strongly dependent on the uncertain value of the spike index $\gamma_{\rm sp}$. Although there are various theoretical studies on the spike index \citep{Gondolo,Gnedin,Merritt}, no promising value of $\gamma_{\rm sp}$ has been obtained from observations. A recent study suggests that using the orbital precession data of the S2 star near Sgr A* might be able to get some constraints on $\gamma_{\rm sp}$ \citep{Chan6}. The more precision future observational data of the S2 star obtained from the GRAVITY collaboration \citep{Abuter} would provide more information on the value of $\gamma_{\rm sp}$. This can help break the degeneracy and provide more concrete constraints on the dark matter parameters.

Another advantage of using the G2 cloud data to constrain annihilating dark matter is that the G2 cloud is very close to the supermassive black hole Sgr A*. The high velocity dispersion of dark matter nearby would enhance the annihilation cross section if the dark matter undergoes p-wave annihilation \citep{Zhao}. This can give very stringent constraints on the velocity-dependent dark matter model based on the forbidden parameter space obtained in Fig.~4. Further studies following our proposed new direction would be required to understand the nature of dark matter.  

\begin{table}
\caption{Parameters of the dark matter density spike model \citep{Fields}}
\begin{tabular}{ |l|l|}
 \hline\hline
$r_{\rm h}$&1.7 pc\\
$r_{\rm b}$&$0.2r_{\rm h}$ (=0.34 pc)\\
$\rho_{\rm b}$& $573\, {\rm M}_{\odot}$pc$^{-3}$ \\
 $t_{\rm ann}$&$10^{10}$ yrs\\
 $\gamma_{\rm in}$&0.5\\
 $r_{\rm in}$& $3.1\times10^{-3}$ pc\\
  \hline\hline
\end{tabular}
\end{table}

\begin{table}
\caption{Minimum allowed dark matter mass such that the heating rate is less than the cooling rate, assuming the spike index $\gamma_{\rm sp}=7/3$ and the thermal annihilation cross section $\langle\sigma v\rangle=2.2\times10^{-26}~{\rm cm^3s^{-1}}$.}
\begin{tabular}{ |l|c|}
 \hline\hline
Channel          & Minimum $m_{\rm DM}$ (GeV) \\
\hline
$\mu^{\pm}$       &40\\
$\tau^{\pm}$ &100\\
$b\bar{b}$     &$\sim 3500$\\
$W^{\pm}$       &$\sim 2000$ \\
 \hline\hline
\end{tabular}
\end{table}

\begin{figure}
\vskip 3mm
\includegraphics[width=70mm]{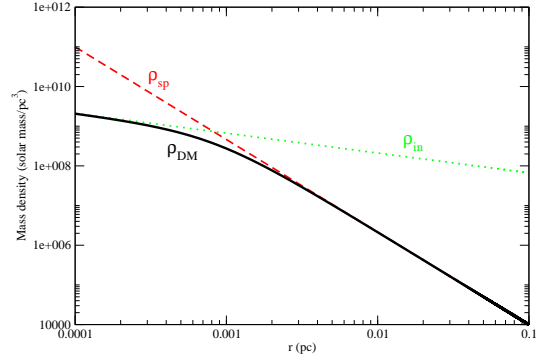}
\caption{The dark matter density spike model (the black solid line). Here, we have taken the spike index $\gamma_{\rm sp}=7/3$, the thermal annihilation cross section $\langle\sigma v\rangle=2.2\times10^{-26}~{\rm cm^3~s^{-1}}$, and $m_{\rm DM}=100$ GeV.}
\label{Fig1}
\end{figure}

\begin{figure}
\vskip 3mm
\includegraphics[width=70mm]{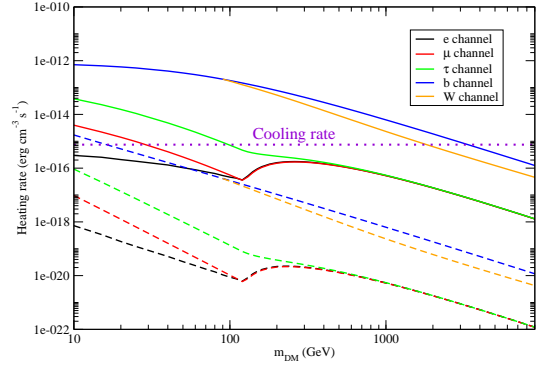}
\caption{Heating rate of the G2 cloud for different dark matter mass and annihilation channels (solid lines: $\gamma_{\rm sp}=7/3$, dashed lines: $\gamma_{\rm sp}=1.5$). Assuming the thermal annihilation cross section $\langle \sigma v \rangle=2.2 \times 10^{-26}$ cm$^3$ s$^{-1}$. The violet dotted line indicates the cooling rate of the G2 cloud.}
\label{Fig2}
\end{figure}

\begin{figure}
\vskip 3mm
\includegraphics[width=70mm]{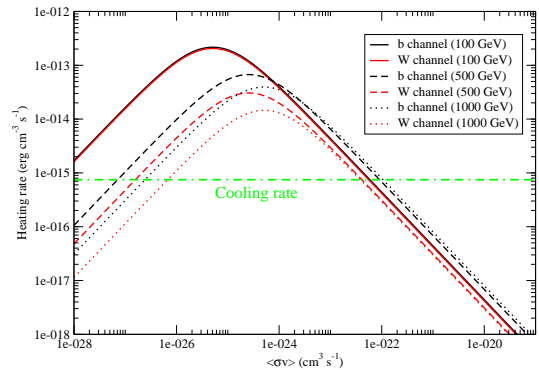}
\caption{The heating rate against annihilation cross section for different dark matter mass and annihilation channels. The green dash-dotted line indicates the cooling rate of the G2 cloud. Here $\gamma_{\rm sp}=7/3$ is assumed.}
\label{Fig3}
\end{figure}

\begin{figure}
\vskip 3mm
\includegraphics[width=80mm]{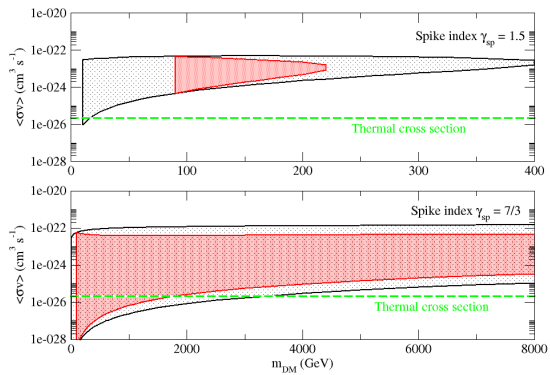}
\caption{The shaded regions indicate the ruled out parameter space of dark matter mass and the annihilation cross section (black: $b\bar{b}$ channel, red: $W^{\pm}$ channel). The green dashed lines indicate the thermal annihilation cross section $\langle \sigma v \rangle=2.2 \times 10^{-26}$ cm$^3$ s$^{-1}$.}
\label{Fig4}
\end{figure}

\section{acknowledgements}
We thank the anonymous referee for useful constructive feedback and comments. The work described in this paper was partially supported by a grant from the Research Grants Council of the Hong Kong Special Administrative Region, China (Project No. EdUHK 18300922).

\section{Data availability statement}
The data underlying this article will be shared on reasonable request to the corresponding author.

\label{lastpage}

\end{document}